\documentclass[12pt]{article}

\begin{document}

\title{\bf \Large Elliptic integral evaluation of a Bessel
moment by contour integration of a lattice Green function}

\author{David Broadhurst\thanks{Department of Physics and Astronomy,
The Open University, Milton Keynes, MK7 6AA, United Kingdom,
{\tt D.Broadhurst@open.ac.uk}.}}

\date{\today}

\maketitle

\abstract{A proof is found for the elliptic integral evaluation
of the Bessel moment $$M:=\int_0^\infty
t\,I_0^2(t)K_0^2(t)K_0(2t)\,{\rm d}t =\frac{1}{12}\,{\bf
K}(\sin(\pi/12)){\bf K}(\cos(\pi/12))
=\frac{\Gamma^6\!\left(\frac13\right)}{64\pi^22^{2/3}}$$
resulting from an angular average of a 2-loop 4-point massive
Feynman diagram, with one internal mass doubled. This evaluation
follows from contour integration of the Green function for a
hexagonal lattice, thereby relating $M$ to a linear combination
of two more tractable moments, one given by the Green function
for a diamond lattice and both evaluated by using W.N.\ Bailey's
reduction of an Appell double series to a product of elliptic
integrals. Cubic and sesquiplicate modular transformations of an
elliptic integral from the equal-mass Dalitz plot are proven and
used extensively. Derivations are given of the sum rules
$$\int_0^\infty\left(I_0(a t)K_0(a t)-\frac{2}{\pi}\,K_0(4a t)
K_0(t)\right)K_0(t)\,{\rm d}t=0$$ with $a>0$, proven by
analytic continuation of an identity from Bailey's work, and
$$\int_0^\infty t\,I_0(a t)\left(I_0^3(a t)K_0(8t)-
\frac{1}{4\pi^2} \,I_0(t)K_0^3(t)\right)\,{\rm d}t=0$$ with $2\ge
a\ge0$, proven by showing that a Feynman diagram in two spacetime
dimensions generates the enumeration of staircase polygons in
four dimensions.}

\newpage

\section{Introduction}

In a recent joint work with David Bailey, Jonathan Borwein and Larry
Glasser~\cite{B3G}
we conjectured, on the basis of numerical computation,
that the moment
\begin{equation}
M:=\int_0^\infty t\,I_0^2(t)K_0^2(t)K_0(2t)\,{\rm d}t
\label{def-M}
\end{equation}
has the evaluation
\begin{equation}
M\stackrel{?}{=}\frac{1}{12}\,K_3K_3^\prime=\frac{\sqrt3}{12}\,K_3^2
=\frac{\Gamma^6\!\left(\frac13\right)}{64\pi^22^{2/3}}
\label{eval-M}
\end{equation}
where $I_0$ and $K_0$ are Bessel functions
and the complete elliptic integral $K_3={\bf K}(\sin(\pi/12))$
is evaluated at the third singular value~\cite{agm},
where the complementary elliptic integral
$K_3^\prime={\bf K}(\cos(\pi/12))$ gives $K_3^\prime/K_3=\sqrt3$
and hence a nome~\cite[17.3.7]{AandS} $q=\exp(-\pi\sqrt3)$.

The moment $M$ is obtained by an angular average, in two-dimensional
Euclidean momentum space, of a 2-loop 4-point
Feynman diagram obtained by cutting two of the 5 lines in a
4-loop vacuum diagram. The arguments of the Bessel functions show
that the external particles have the same mass as two of the
three internal particles, while the third internal mass is doubled.
In~\cite[Sect.~5.10]{B3G}, a proof was found for Laporta's conjectural
evaluation~\cite{Lap21} of the equal-mass moment\footnote{Some equal-mass
moments with only 4 Bessel functions
evaluate~\cite{ising,groote,ouvry} in terms of 
$\zeta(2)$ and $\zeta(3)$.}
$\int_0^\infty t\,I_0^2(t)K_0^3(t)\,{\rm d}t$ at the 15th singular
value with the much smaller nome
$q=\exp(-\pi\sqrt{15})\approx5.2\times10^{-6}$.
Notwithstanding the substantial progress made in~\cite{B3G} since my talk
{\sl Reciprocal PSLQ and the Tiny Nome of Bologna} at
Bielefeld\footnote{See {\tt
http://www.physik.uni-bielefeld.de/igs/schools/ZiF2007/Broadhurst.pdf}~.}
in June 2007, we were unable to prove that doubling one of the
\emph{internal} masses leads to the nome $q=\exp(-\pi\sqrt{3})$, though
it was possible to prove that this nome results from doubling one
of the external masses.

Alternative forms of conjecture~(\ref{eval-M}) were given
in~\cite{B3G}. These correspond to the proven evaluations
\begin{eqnarray}
M&=&\frac14
\int_0^\infty\int_0^\infty\frac{
\,{\rm d}x\,{\rm d}y}{\sqrt{(1+x^2)(1+y^2)(1+(x+y)^2)(1+(x-y)^2)}}
\label{double-M}\\
&=&\frac14
\int_0^1{\bf K}\left(\frac{z\sqrt{5-4z}}{2-z}\right)\frac{{\rm d}z}{2-z}
\label{planar-M}\\
&=&
\int_0^1\frac{y\,{\bf K}\left(
\frac{2+y}{2-y}\sqrt{\frac{1-y}{1+y}}\right)}
{\sqrt{(2-y)^3(2+y)(1+y)}}\,{\rm d}y\,.
\label{disp-M}
\end{eqnarray}
It was also proven that each of the integrals
\begin{eqnarray}
\int_0^\infty\int_0^\infty\frac{{\rm d}x\,{\rm d}y}
{\sqrt{(1+x^2)(1+y^2)(1+(x+y)^2)}}
&=&\frac{2}{3}\,K_3K_3^\prime
\label{third-plus}\\
\int_0^\infty\int_0^\infty\frac{{\rm d}x\,{\rm d}y}
{\sqrt{(1+x^2)(1+y^2)(1+(x-y)^2)}}
&=&\frac{4}{3}\,K_3K_3^\prime
\label{third-minus}\\
\int_0^1\frac{{\bf K}(y)\,{\rm d}y}
{\sqrt{(1-y^2)(1+3y^2)}}&=&\frac12\,K_3K_3^\prime
\label{third-honolulu}
\end{eqnarray}
yields the third singular value.
Yet no combination of these 6 interesting formulae
appeared to offer a direct route to proving
the outstanding conjecture that $M=\frac{1}{12}\,K_3K_3^\prime$.

I have now found a proof by a route that is very
far from direct, namely by
using the vanishing of a suitable contour integral of the
Green function~\cite[Eq.~4.7]{Joyce94}
\begin{equation}
\widetilde{D}(z):=
\frac{1}{\pi^2}\int_0^\pi\int_0^\pi
\frac{{\rm d}\theta_1\,{\rm d}\theta_2}
{1-z^2(3+2\cos\theta_1+4\cos\theta_1\cos3\theta_2)}
\label{hex}
\end{equation}
for a two-dimensional hexagonal (or ``honeycomb") lattice.
Then, after many intermediate transformations, I am able
to show that $M$ is the \emph{difference} of an even and odd
Bessel moment, each of which are now proven to yield the third
singular value.

In Section~2, I use contour integration to derive the vanishing of
\begin{equation}
\int_\frac13^1\sigma_1(x)\,{\rm d}x+
\int_1^\infty\sigma_2(x)\,{\rm d}x-
\int_0^\infty\widetilde{D}({\rm i}x)\,{\rm d}x=0
\label{contour}
\end{equation}
where $\sigma_1$ and $\sigma_2$
are the reciprocals of arithmetic-geometric means (AGMs).

In Section~3, I evaluate the integral of
$\widetilde{D}$ on the imaginary axis
as an \emph{even} moment of 3 Bessel functions.
By analytic continuation of results
obtained by Wilfrid Norman Bailey~\cite{bailey1936a,bailey1936},
I then obtain the third singular value for this term in the
contour integral~(\ref{contour}).
Moreover, I prove the remarkable result that for arbitrary
real positive $a$
\begin{equation}
{\mathcal K}_0(a,t):=I_0(a t)K_0(a t)-\frac{2}{\pi}\,K_0(4a t)K_0(t)
\label{Kat}
\end{equation}
is \emph{orthogonal} to $K_0(t)$, giving a continuous infinity
of sum rules of the form
\begin{equation}
\int_0^\infty {\mathcal K}_0(a,t)K_0(t)\,{\rm d}t=0\,.
\label{sum-rule}
\end{equation}

In Section~4, I derive a cubic modular transformation
that relates the integral of $\sigma_2$
to the integral of $\widetilde{D}$
between the origin and the branchpoint at $z=\frac13$.
I also give a simple proof of a result recorded in~\cite[Eq.~4.6.14]{agm}
and used in~\cite[Sect.~5.10]{B3G} to derive a modular identity from
a transformation in~\cite{davdel} of the equal mass Dalitz plot.

In Section~5, I derive a sesquiplicate modular transformation
of the form $q\to q^{3/2}$.

In Section~6, I combine these two modular transformations
to relate the integral of $\sigma_2$
to an \emph{odd} moment of 5 Bessel functions that yields
the third singular value via its relation to a Green
function on a three-dimensional diamond lattice.

In Section~7, I prove that $M=\frac{1}{12}\,K_3K_3^\prime$.

In Section~8, I derive further identities, by combining
cubic and sesquiplicate modular transformations.

\section{Spectral function and contour integral}

The Taylor series~\cite[Eq.~184]{B3G}
\begin{equation}
\widetilde{D}(z)=\sum_{k=0}^\infty a_k z^{2k}
\label{exp-Dt}
\end{equation}
is valid for $|z|<\frac13$ and has integer coefficients\footnote{See {\tt
http://www.research.att.com/\~{ }njas/sequences/A002893}~.}
\begin{equation}
a_k=\sum_{j=0}^k{k\choose j}^2{2j\choose j}
\label{a-hexagonal}
\end{equation}
enumerating closed walks with $2k$ steps on a hexagonal lattice.
For $\frac13>y>-\frac13$, there is an exponentially fast method
for computing~\cite[Eq.~4.10]{Joyce94}
\begin{equation}
\widetilde{D}(y)=
\frac{1}{{\rm AGM}(\sqrt{(1-3y)(1+y)^3},\sqrt{(1+3y)(1-y)^3})}
\label{AGM-Dt}
\end{equation}
as the reciprocal of an arithmetic-geometric mean.

I now give the analytic continuation of~(\ref{AGM-Dt}) in
the complex $z$-place with cuts on the real axis running from
$z=-\infty$ to $z=-\frac13$ and from $z=\frac13$ to $z=+\infty$.
To construct $\widetilde{D}$, I use the dispersion relation~\cite{Gabriel}
\begin{equation}
\widetilde{D}(z)=
\frac{2}{\pi}\int_\frac13^1
\frac{\sigma_1(x)x}{x^2-z^2}\,{\rm d}x+
\frac{2}{\pi}\int_1^\infty
\frac{\sigma_2(x)x}{x^2-z^2}\,{\rm d}x
\label{Dt-disp}
\end{equation}
for which it suffices to know the
spectral function~\cite[Chap.~6]{Gabriel}
$\sigma(x)=\Im\widetilde{D}(x+{\rm i}\varepsilon)$,
which is the imaginary part on the top lip of the right-hand cut,
with real $x>\frac13$ and infinitesimal real $\varepsilon>0$.
Denoting $\sigma(x)=\sigma_1(x)$, for $\frac13<x<1$, and
$\sigma(x)=\sigma_2(x)$, for $1<x$, I obtain
\begin{eqnarray}
\sigma_1(x)&=&
\frac{1}{{\rm AGM}(\sqrt{16x^3},\sqrt{(1+3x)(1-x)^3})}
\label{sigma1}\\
\sigma_2(x)&=&
\frac{1}{{\rm AGM}(\sqrt{(3x-1)(x+1)^3},\sqrt{(3x+1)(x-1)^3})}
\label{sigma2}
\end{eqnarray}
from analytic continuation of~(\ref{AGM-Dt}).

Thus moments of these reciprocal AGMs yield the Taylor coefficients
\begin{equation}
\frac{2}{\pi}\int_\frac13^1
\frac{\sigma_1(x)}{x^{2k+1}}\,{\rm d}x
+\frac{2}{\pi}\int_1^\infty
\frac{\sigma_2(x)}{x^{2k+1}}\,{\rm d}x
=\sum_{j=0}^k{k\choose j}^2{2j\choose j}
\label{a-sigma}
\end{equation}
as was confirmed by \emph{Pari-GP}, which yielded 200 good decimal digits
for each of the first 100 Taylor coefficients in 73 CPU-seconds.

To prove that $M=\frac{1}{12}\,K_3K_3^\prime$ it will
be sufficient to take the imaginary part of the contour integral
\begin{equation}
\oint_C \widetilde{D}(z)\,{\rm d}z=
\int_0^\frac13\widetilde{D}(y)\,{\rm d}y
+\int_\frac13^\infty\widetilde{D}(x+{\rm i}\varepsilon)\,{\rm d}x
-{\rm i}\int_0^\infty\widetilde{D}({\rm i}x)\,{\rm d}x=0
\label{contour-both}
\end{equation}
where $C$ is a counterclockwise contour that encloses
the quadrant in which the real and imaginary parts
of $z$ are positive. Cauchy's theorem ensures
the vanishing of~(\ref{contour-both}), since the quarter circle
at infinity gives no contribution. Indeed, large $x$ behaviour
\begin{equation}
\sqrt3\,\sigma_2(x)=\frac{1}{x^2}
+{\rm O}\left(\frac{1}{x^4}\right)
\label{s2-asy}
\end{equation}
shows that $\pi\sqrt3\,\widetilde{D}(z)=\log(-z^2)/z^2+{\rm O}(1/z^2)$.
More precisely, I obtained from~\cite[Sect.~5.7]{B3G} the large $z$
behaviour
\begin{equation}
\pi\sqrt3\,\widetilde{D}(z)=
-\frac{\log(-9z^2)}{z^2}+{\rm O}\left(\frac{\log(-9z^2)}{z^4}\right)\,.
\label{Dt-asy}
\end{equation}

Taking the imaginary part of~(\ref{contour-both}), I obtain
the sum rule~(\ref{contour}) and commence the contour clockwise,
from the origin, starting on the imaginary axis.

\section{An even Bessel moment on the imaginary axis}

{}From~\cite[Eq.~23]{B3G} I obtain the odd moments
\begin{equation}
\int_0^\infty t^{2k+1}I_0(t)K_0^2(t)\,{\rm d}t
=\frac{\pi}{3\sqrt3}\left(\frac{2^k{}k!}{3^k}\right)^2a_k
\label{odd-moments}
\end{equation}
and hence for $\frac13>x>-\frac13$ the representation
\begin{equation}
\widetilde{D}(x)=\frac{3\sqrt3}{\pi}
\int_0^\infty t\,I_0(3x t)I_0(t)K_0^2(t)\,{\rm d}t
\label{Dt-moment}
\end{equation}
using the expansion~\cite[9.6.12]{AandS}
\begin{equation}
I_0(z)=\sum_{k=0}^\infty\left(\frac{z^k}{2^k{}k!}\right)^2\,.
\label{I0}
\end{equation}
The analytic continuation to the imaginary axis is
\begin{equation}
\widetilde{D}({\rm i}x)=\frac{3\sqrt3}{\pi}\,E(3x)
\label{Dt-E}
\end{equation}
with the Bessel moment
\begin{equation}
E(x)=\int_0^\infty t\,J_0(x t)I_0(t)K_0^2(t)\,{\rm d}t
\label{E-moment}
\end{equation}
obtained from~(\ref{Dt-moment}) and~(\ref{Dt-E}) by
the analytic continuation $J_0(x t)=I_0({\rm i}x t)$. Its
evaluation
\begin{equation}
E(x)=\frac{\pi}{{\rm AGM}(\Re h,|h|)}\,,\quad\mbox{ with }\quad
h=\sqrt{(1-{\rm i}x)(3+{\rm i}x)^3}\,,
\label{E-h}
\end{equation}
is obtained by substituting $y={\rm i}x/3$ in~(\ref{AGM-Dt})
and by using ${\rm AGM}(h,\overline{h})={\rm AGM}(\Re h,|h|)$,
where $\overline{h}$ is the complex conjugate of $h$.
I remark that~(\ref{E-h}) is a computationally efficient rewriting
of~\cite[Eq.~138]{B3G}, since the former requires only a complex square root,
while the latter used a less frugal, but equivalent,
combination of trigonometric and inverse trigonometric functions.

Using~(\ref{E-moment}), I obtain
\begin{equation}
\int_0^\infty E(x)\,{\rm d}x=\int_0^\infty I_0(t)K_0^2(t)\,{\rm d}t
\label{E-int}
\end{equation}
as an even moment of Bessel functions, by interchanging
the order of integrations over $x$ and $t$ and using the
evaluation~\cite[11.4.17]{AandS} $\int_0^\infty t\,J_0(x t)\,{\rm d}x=1$.

Thanks to~\cite{bailey1936a,bailey1936}, I was
able to obtain the remarkable identity
\begin{equation}
\int_0^\infty I_0(a t)K_0(a t)K_0(t)\,{\rm d}t
=\frac{2}{\pi}\int_0^\infty K_0(4a t)K_0^2(t)\,{\rm d}t
\label{remarkable}
\end{equation}
for all real $a>0$. I proved this by using~\cite[Eq.~3.3]{bailey1936}
in the case $0<a\le\frac12$, obtaining the left-hand side of~(\ref{remarkable})
as a product of complete elliptic integrals that is identical
to the product that I had earlier obtained by the delicate limiting
process that led to~\cite[Eq.~36]{B3G}, for the evaluation of the
more demanding right-hand side. Then, for $a\ge\frac12$, the analytic
continuation given in~\cite[Eqs.~34,35]{B3G}
provides the evaluation
\begin{equation}
4a\int_0^\infty I_0(a t)K_0(a t)K_0(t)\,{\rm d}t=
{\bf K}^2(\sin\alpha)+{\bf K}^2(\cos\alpha)\,,
\quad\alpha=\frac12\,\arcsin(\frac{1}{2a})
\label{K-alpha}
\end{equation}
and hence the third singular value in the neat evaluation
\begin{equation}
\int_0^\infty E(x)\,{\rm d}x=\int_0^\infty I_0(t)K_0^2(t)\,{\rm d}t
=K_3^2
\label{neat-E}
\end{equation}
obtained by setting $a=1$ and hence $\alpha=\pi/12$ in~(\ref{K-alpha}).

Thus I obtain
\begin{equation}
\int_\frac13^1\sigma_1(x)\,{\rm d}x+
\int_1^\infty\sigma_2(x)\,{\rm d}x-
\frac{1}{\pi}\,K_3K_3^\prime=0
\label{contour2}
\end{equation}
from~(\ref{contour}), by using~(\ref{Dt-E}) and the relation
$K_3^\prime=\sqrt3K_3$.

\section{A cubic modular transformation}

Proceeding clockwise, I pass along the quarter circle at infinity
with impunity, thanks to~(\ref{Dt-asy}), and arrive at the
top lip of the cut, with $z=x+{\rm i}\varepsilon$ and large $x$, where
the spectral function $\sigma_2(x)$ is needed.
Here I shall substitute $3y=1/x$ in the identity
\begin{eqnarray}
&&\frac{1}{{\rm AGM}(\sqrt{(1-3y)(1+y)^3},\sqrt{(1+3y)(1-y)^3})}
\nonumber\\
&=&\frac{1}{{\rm AGM}(\sqrt{(1-y)(1+3y)^3},\sqrt{(1+y)(1-3y)^3})}
\label{cubic-mod}
\end{eqnarray}
which is valid for $\frac13>y>-\frac13$
and was proven
using the {\tt EllipticK} and {\tt HeunG} functions
of \emph{Maple} to show that each side of~(\ref{cubic-mod})
satisfies the same second-order differential as
${\rm HeunG}(9,3;1,1,1,1;9y^2)$. Since each side
has the Taylor expansion $1+3y^2+{\rm O}(y^4)$,
each evaluates to this {\tt HeunG} function, for $\frac13>y>-\frac13$.

I remark that~(\ref{cubic-mod}) is a cubic modular
transformation. Setting
\begin{equation}
k=\sqrt{\frac{16y^3}{(1+3y)(1-y)^3}}\,,\quad k^\prime=\sqrt{1-k^2}\,,\quad
l=\sqrt{\frac{16y}{(1-y)(1+3y)^3}}\,,\quad l^\prime=\sqrt{1-l^2}\,,
\label{k-and-l}
\end{equation}
one trivially obtains
\begin{equation}
\sqrt{k^\prime l^\prime}+\sqrt{k l}=
\frac{(1+y)(1-3y)+4y}{(1-y)(1+3y)}=1
\label{proven-mod}
\end{equation}
for $\frac13>y>-\frac13$. This proves~\cite{agm}
that the modular transformation is cubic, i.e.\ that
the nome~\cite[17.3.7]{AandS}
associated with $k$ is the cube of the nome associated with $l$.
Denoting the latter by
$q=\exp(-\pi\,{\bf K}(l^\prime)/{\bf K}(l))$, one may
rewrite~(\ref{cubic-mod}) as
\begin{equation}
\frac{\theta_2^2(q^3)}{\sqrt{16y^3}}=\frac{\theta_2^2(q)}{\sqrt{16y}}
\label{J-C}
\end{equation}
with $\theta_2(q)=\sum_{n=-\infty}^\infty q^{(n+1/2)^2}$.
Then $\theta_3(q)=\sum_{n=-\infty}^\infty q^{n^2}$ determines
$l^2=\theta_2^4(q)/\theta_3^4(q)$. Hence I have proven that
the cubic multiplier $y=\theta_2^2(q^3)/\theta_2^2(q)$ is a root
of the quartic polynomial
$(1-y)(1+3y)^3-16y\theta_3^4(q)/\theta_2^4(q)$.
Thus I recover a result
of Joubert and Cayley that was recorded in~\cite[Eq.~4.6.14]{agm}
and used in the proof of the modular identity in~\cite[Eq.~157]{B3G},
obtained from an analysis of the equal mass Dalitz plot in~\cite{davdel}.

More importantly, for present considerations, one may
rewrite~(\ref{cubic-mod}) as
\begin{equation}
\widetilde{D}(y)=\sqrt3x^2\sigma_2(x)\,,\quad\mbox{ for }\quad 3xy=1\,,
\label{s2-Dt}
\end{equation}
and hence obtain
\begin{equation}
\int_0^\frac13\widetilde{D}(y)\,{\rm d}y=
\frac1{\sqrt3}\int_1^\infty\sigma_2(x)\,{\rm d}x
\end{equation}
which transforms~(\ref{contour2}) to
\begin{equation}
\int_\frac13^1\sigma_1(x)\,{\rm d}x+
\sqrt3\int_0^\frac13\widetilde{D}(y)\,{\rm d}y-
\frac{1}{\pi}\,K_3K_3^\prime=0\,.
\label{contour3}
\end{equation}

\section{A sesquiplicate modular transformation}

The integral over $\widetilde{D}$ in~(\ref{contour3})
may be transformed to yield a Bessel moment
by using the transformation
\begin{equation}
2\widetilde{D}(y)=\sqrt3(1-x^2)\sigma_1(x)\,,
\quad\mbox{ for }\quad 9(1-x^2)(1-y^2)=8\,,
\label{a-to-b}
\end{equation}
which is valid for $0<x<1$.
It maps the region $0<x<\frac13$ to $\frac13>y>0$
and the region $\frac13<x<1$ to the positive imaginary $y$-axis.
To prove~(\ref{a-to-b}), I used the representations
\begin{equation}
\widetilde{D}(y)={\rm HeunG}(9,3;1,1,1,1;9y^2)\,,\quad
\frac{4\sigma_1(x)}{3\sqrt3}={\rm HeunG}(-8,-2;1,1,1,1;1-9x^2)\,,
\label{E-HeunG}
\end{equation}
with the latter proven in the same manner as was used
to proved the former, in Section~4.
Applying the penultimate identity of \emph{Maple}'s
{\tt FunctionAdvisor(identities,HeunG)}
to $\sigma_1(x)$, one obtains~(\ref{a-to-b}), provided that one avoids
the cut with $9y^2\ge1$.
This is a modular transformation of the form $q\to q^{3/2}$,
with the nomes ``in sesquiplicate proportion" (as Newton's
translator Motte said of the corresponding
power law relation in Kepler's third law).
It may be obtained by combining an ascending cubic modular
transformation with a descending quadratic modular transformation.
Thanks to the explicit form of Heun's differential equation~\cite{heun}
I was able to find a more direct proof, by using~(\ref{E-HeunG}).

Applying~(\ref{a-to-b}) in the region $0<y<\frac13$, I obtain
\begin{equation}
\frac{\pi\sqrt3}{2}\int_0^\frac13\widetilde{D}(y)\,{\rm d}y
=\int_0^\frac13\frac{D(x)}{\sqrt{(1-x^2)(1-9x^2)}}\,{\rm d}x
\label{second-trans}
\end{equation}
where the square root comes from the Jacobian
\begin{equation}
\left|\frac{{\rm d}y}{{\rm d}x}\right|=\frac{8x}{3(1-x^2)}\,
\frac{1}{\sqrt{(1-x^2)(1-9x^2)}}
\label{Jacobian}
\end{equation}
of the transformation in~(\ref{a-to-b}) and
\begin{equation}
D(x):=2\pi x\sigma_1(x)
=\frac{4x\,{\bf K}\left(\sqrt{\frac{(1-3x)(1+x)^3}{(1+3x)(1-x)^3}}\right)}
{\sqrt{(1+3x)(1-x)^3}}
\label{Dy}
\end{equation}
is the function defined in~\cite[Eq.~64]{B3G}.
I remark that $D$
provides an evaluation of the moment~\cite[Eq.~149]{B3G}
\begin{equation}
\int_0^\infty t\,I_0^2(t)K_0(t)K_0(ct)\,{\rm d}t
=\frac{1}{6c}\,D\left(\frac{1}{c}\right)
\label{Dy-moment}
\end{equation}
for $c>1$. The physical significance of $D$
derives from the dispersion relation~\cite[Sect.~4.2]{B3G}
for an odd moment with 4 Bessel functions,
\begin{equation}
S(w^2):=\int_0^\infty t\,J_0(w t)K_0^3(t)\,{\rm d}t=
\int_0^\frac13\frac{D(x)\,{\rm d}x}{1+w^2x^2}\,,
\label{Dy-disp}
\end{equation}
which is the sunrise diagram~\cite{groote,sunrise} with 3 unit internal masses
and Euclidean external momentum with norm $w^2$, in
two-dimensional spacetime. On the cut with $-w^2=c^2>9$,
the elliptic integral in $D(1/c)$ appears in the imaginary part
of $S$ from integration over the Dalitz plot~\cite{davdel} for the
decay of a particle of mass $c>3$ into 3 particles of unit mass.

Then the 5-Bessel moment
\begin{equation}
T(u^2,v^2):=\int_0^\infty t\,J_0(u t)J_0(v t)K_0^3(t)\,{\rm d}t
=\frac{1}{\pi}\int_0^\pi S(u^2+2uv\cos\theta+v^2)
\label{T5}
\end{equation}
is obtained as an angular average~\cite[Sect.~5.3]{B3G} of a diagram
in which the external momentum is shared by a pair
of particles.  Exchanging the
order of integration in~(\ref{Dy-disp}) and~(\ref{T5}), I obtain
\begin{equation}
T(u^2,v^2)=\int_0^\frac13\frac{D(x)\,{\rm d}x}
{\sqrt{(1+(u x-v x)^2)(1+(u x+v x)^2)}}
\label{T5-D}
\end{equation}
which may be analytically continued to negative values of $u^2$ and $v^2$,
provided that the argument $1+2(u^2+v^2)x^2+(u^2-v^2)^2x^4$ of the
square root remains positive in the integration region $1>9x^2>0$.
Hence I prove that
\begin{equation}
T(-a^2,-1)=\int_0^\infty t\,I_0(a t)I_0(t)K_0^3(t)\,{\rm d}t
=\int_0^\frac13\frac{D(x)\,{\rm d}x}
{\sqrt{(1-(a x-x)^2)(1-(a x+x)^2)}}
\label{W-D}
\end{equation}
for $2\ge a\ge0$

\section{Bessel moments from the diamond lattice}

In~\cite[Eq.~55]{B3G}, the Bessel moments
\begin{equation}
\int_0^\infty t^{2k+1}I_0(t)K_0^3(t)\,{\rm d}t
=\frac{\pi^2}{16}\left(\frac{k!}{4^k}\right)^2b_k
\label{odd-moments-b}
\end{equation}
were evaluated in terms of the integers\footnote{See {\tt
http://www.research.att.com/\~{ }njas/sequences/A002895}~.}
\begin{equation}
b_k=\sum_{j=0}^k{k\choose j}^2{2k-2j\choose k-j}{2j\choose j}
\label{diamond-b}
\end{equation}
that enumerate closed walks on a diamond lattice.
They give the Taylor coefficients of
\begin{equation}
\int_0^\infty t\,I_0(a t)I_0(t)K_0^3(t)\,{\rm d}t
=\frac{\pi^2}{16}\sum_{k=0}^\infty b_k \left(\frac{a}{8}\right)^{2k}
\label{diamond-moment}
\end{equation}
as may seen by expanding $I_0(a t)$ and using~(\ref{odd-moments-b}).
It follows that this Bessel moment is given by the
evaluation~\cite[Eq.~5.4]{Joyce94} of the diamond lattice Green function,
\begin{equation}
\int_0^\infty t\,I_0(a t)I_0(t)K_0^3(t)\,{\rm d}t
=\frac14\,{\bf K}(k_-)\,{\bf K}(k_+),
\label{diamond-green}
\end{equation}
where the arguments of the complete elliptic integrals are determined by
\begin{equation}
k_\pm^2=\frac12\pm\frac{a^2}{8}\sqrt{1-\frac{a^2}{16}}
-\left(\frac12-\frac{a^2}{16}\right)\sqrt{1-\frac{a^2}{4}}\
\label{k2}
\end{equation}
by setting $z=a/2$ in~\cite[Eq.~5.5]{Joyce94}.
Once again, a cubic modular transformation is involved,
since $\sqrt{k_-^\prime k_+^\prime}+\sqrt{k_-k_+}=1$,
with $k_\pm^\prime=\sqrt{1-k_\pm^2}$.

Then setting $a=2$ in~(\ref{W-D}) and~(\ref{diamond-green}),
I obtain an evaluation of the integrals~(\ref{second-trans})
\begin{equation}
\frac{\pi\sqrt3}{2}\int_0^\frac13\widetilde{D}(y)\,{\rm d}y=
\int_0^\frac13\frac{D(x)\,{\rm d}x}
{\sqrt{(1-x^2)(1-9x^2)}}=\frac14\,K_3K_3^\prime
\label{W-D-2}
\end{equation}
with the third singular value $k_3=\frac12\sqrt{2-\sqrt3}=\sin(\pi/12)$
resulting from $k_-$ in~(\ref{k2}) at $a=2$. Hence
I complete the evaluation of the middle term in the contour
integral~(\ref{contour3}) and obtain the pleasing result
\begin{equation}
\int_\frac13^1\sigma_1(x)\,{\rm d}x+
\frac{1}{2\pi}\,K_3K_3^\prime-
\frac{1}{\pi}\,K_3K_3^\prime=0
\label{contour4}
\end{equation}
which shows that the first term also has an evaluation at the third
singular value.

I remark that the diamond lattice integers~(\ref{diamond-b})
also enumerate staircase polygons~\cite{glasser,guttmann}
in \emph{four} dimensions, for which the generating
function~\cite[Eq.~6a]{glasser} is an odd 5-Bessel moment
containing $I_0^4$. Thus I derive from~(\ref{diamond-green})
the remarkable sum rule
\begin{equation}
\int_0^\infty t\,I_0(a t)\left(I_0^3(a t)K_0(8t)
-\frac{1}{4\pi^2}\,I_0(t)K_0^3(t)\right)\,{\rm d}t=0
\label{remarkable5}
\end{equation}
for $2\ge a\ge0$.

\section{Proof of the conjecture}

Now I set $y={\rm i}w/3$ in the sesquiplicate modular
transformation~(\ref{a-to-b})
and obtain an integral of $\widetilde{D}$ on the imaginary
axis. The result is
\begin{equation}
\frac{\pi}{6}\int_\frac13^1\sigma_1(x)\,{\rm d}x=
\int_0^\infty\frac{E(w)w}{\sqrt{(w^2+1)(w^2+9)}}\,{\rm d}w
\label{Jacob}
\end{equation}
with a square root from the inverse of the Jacobian~(\ref{Jacobian}) and
\begin{equation}
E(w)=\int_0^\infty t\,J_0(w t)I_0(t)K_0^2(t)\,{\rm d}t
\label{E-moment-w}
\end{equation}
coming from the analytic continuation~(\ref{Dt-E}).

Then I observe that
\begin{eqnarray}
\int_0^\infty s\,J_0(w s)J_0(v s)K_0(c s)\,{\rm d}s&=&
\frac{1}{\pi}\int_0^\pi\frac{{\rm d}\theta}
{w^2+2wv\cos\theta+v^2+c^2}
\label{J-J-K}\\
&=&\frac{1}{\sqrt{((w-v)^2+c^2)((w+v)^2+c^2)}}
\label{triangle}
\end{eqnarray}
since the Bessel moment corresponds to an angular
average~\cite[Sect.~5.3]{B3G} of a tree diagram in
two-dimensional Euclidean momentum space.
Using the distribution
\begin{equation}
\int_0^\infty w J_0(w s)J_0(w t)\,{\rm d}w=2\delta(s^2-t^2)
\label{J-J}
\end{equation}
I evaluate the 5-Bessel moment
\begin{equation}
\int_0^\infty t\,J_0(v t)K_0(c t)I_0(t)K_0^2(t)\,{\rm d}t=
\int_0^\infty\frac{E(w)w}{\sqrt{((w-v)^2+c^2)((w+v)^2+c^2)}}\,{\rm d}w
\end{equation}
as a folding of~(\ref{E-moment-w}) with~(\ref{J-J-K}).

Then I set $v={\rm i}$ and $c=2$ and obtain
\begin{equation}
\frac{\pi}{6}\int_\frac13^1\sigma_1(x)\,{\rm d}x=
\int_0^\infty t\,I_0^2(t)K_0^2(t)K_0(2t)\,{\rm d}t:=M
\label{M-at-last}
\end{equation}
from~(\ref{Jacob}). Hence the contour integral~(\ref{contour4}) gives
\begin{equation}
\frac{6}{\pi}\,M+
\frac{1}{2\pi}\,K_3K_3^\prime-
\frac{1}{\pi}\,K_3K_3^\prime=0
\label{contour5}
\end{equation}
which completes the proof that $M=\frac{1}{12}\,K_3K_3^\prime$.

The reader may consider (as does the author) that this is a
rather indirect proof, since it involves the difference
between an even and odd Bessel moment, obtained via delicate contour
integration and several modular transformations. However, a great
deal of unrewarded effort had previously been expended on
searching for a more direct proof.

\section{Further identities}

Now I consider the real part
on the cut, remarking that
\begin{equation}
\rho(x)=
\frac{1}{{\rm AGM}(\sqrt{(3x-1)^3(x+1)},\sqrt{16x})}
\label{rho}
\end{equation}
is real for $x>\frac13$ and satisfies the same differential
equation as $\widetilde{D}$, $\sigma_1$ and $\sigma_2$. Thus a
multiple of $\rho$ should give the real part of $\widetilde{D}$
on each of the two portions of the cut, with $\frac13<x<1$ and $1<x$.
However, these multiples may be distinct, since the two portions
are separated by a singular point of the differential equation,
at $x=1$. Indeed the multiples are not the same;
the Green function on the lip of its cut is given by
\begin{equation}
\widetilde{D}(x\pm{\rm i}\varepsilon)=\left\{\begin{array}{rl}
\rho(x)\pm{\rm i}\sigma_1(x)&\quad\mbox{ for }\quad\frac13<x<1\,,\\
-2\rho(x)\pm{\rm i}\sigma_2(x)&\quad\mbox{ for }\quad1<x\,,
\end{array}\right.
\label{real-part}
\end{equation}
where the unit multiple, on the first portion of the cut,
results from analytic continuation of~(\ref{AGM-Dt}), while the factor
of $-2$, for the second portion, is entirely determined by the
asymptotic behaviour~(\ref{Dt-asy}).

Thus I derive the vanishing of
\begin{equation}
\Re\oint_C\widetilde{D}(z)\,{\rm d}z
=\int_0^\frac13\widetilde{D}(y)\,{\rm d}y
+\int_\frac13^1\rho(x)\,{\rm d}x
-2\int_1^\infty\rho(x)\,{\rm d}x=0
\label{real-contour}
\end{equation}
and use the cubic modular transformation complementary to~(\ref{s2-Dt}),
\begin{equation}
3\sqrt3x^2\rho(x)=\sigma_1(y)\,,\quad\mbox{ for }\quad3xy=1\,,
\label{rho-cubic}
\end{equation}
to write the contour integral as
\begin{equation}
\int_0^\frac13\widetilde{D}(y)\,{\rm d}y
+\frac{1}{\sqrt3}\int_\frac13^1\sigma_1(y)\,{\rm d}y
-\frac{2}{\sqrt3}\int_0^\frac13\sigma_1(y)\,{\rm d}y=0\,.
\label{rho-trans}
\end{equation}
The first term was evaluated in~(\ref{W-D-2})
and the second in~(\ref{contour4}). Hence I prove the identity
\begin{equation}
K_3K_3^\prime=\int_0^\frac13\frac{D(y)}{y}\,{\rm d}y
\label{another}
\end{equation}
obtained by using the definition $D(y):=2\pi y\sigma_1(y)$
in the third term.

There are now four proven evaluations at the third singular value
from integrals of $D$, $\widetilde{D}$ and $E$, namely
\begin{eqnarray}
K_3^2
&=&2\pi\int_0^\frac13\widetilde{D}(y)\,{\rm d}y
=\int_0^\infty E(x)\,{\rm d}x
\label{K-Dt}\\
K_3K_3^\prime
&=&\int_0^\frac13\frac{D(y)}{y}\,{\rm d}y
=\int_\frac13^1\frac{D(y)}{y}\,{\rm d}y
\label{K-D}
\end{eqnarray}
obtained from two Bessel moments and from
the vanishing
of the real and imaginary parts of a contour integral.

There are now three proven evaluations of odd Bessel moments
at the third singular value, namely
\begin{eqnarray}
K_3K_3^\prime
&=&4\int_0^\infty t\,I_0(2t)I_0(t)K_0^3(t)\,{\rm d}t\
\label{external}\\
&=&4\pi^2\int_0^\infty t\,I_0^4(t)K_0(4t)\,{\rm d}t
\label{staircase}\\
&=&12\int_0^\infty t\,I_0^2(t)K_0^2(t)K_0(2t)\,{\rm d}t:=12M
\label{odd-Bessel}
\end{eqnarray}
with the first proven in~\cite{B3G}, the second in~\cite{glasser}
and the third by the contour integration~(\ref{contour5}).

There are now four proven evaluations of even Bessel moments
at the third singular value, namely
\begin{eqnarray}
K_3^2=\int_0^\infty I_0(t)K_0^2(t)\,{\rm d}t
=\frac{2}{\pi}\int_0^\infty K_0^2(t)K(4t)\,{\rm d}t\,,
\label{odd-Bessel-1}\\
K_3K_3^\prime=4\int_0^\infty I_0(t)K_0(t)K_0(4t)\,{\rm d}t
=\frac{2}{\pi}\int_0^\infty K_0^3(t)\,{\rm d}t\,,
\label{odd-Bessel-2}
\end{eqnarray}
of which only the last was proven in~\cite{B3G}.
The first is proven in~(\ref{neat-E}), the second
results from sum rule~(\ref{sum-rule}) at $a=1$
and the third from setting $a=\frac14$ in~(\ref{sum-rule})
and then rescaling $t$ by a factor of 4.

Moreover, by using the Taylor expansion of $\widetilde{D}$
in~(\ref{K-Dt}) and the Clausen product formula~\cite{agm} for
$K_3^2$, I obtain a novel relation
\begin{equation}
\sum_{k=0}^\infty\frac{a_k}{(2k+1)3^{2k+1}}
=\frac{\pi}{8}\sum_{k=0}^\infty\frac{{2k\choose k}^3}{2^{8k}}
\label{sums}
\end{equation}
between a sum over the integers $a_k$, which enumerate
closed walks on a two-dimensional hexagonal lattice,
and a sum over the integers ${2k\choose k}^3$,
which enumerate closed walks on a three-dimensional
body centred cubic lattice~\cite{Joyce94}.

Relations between integrals of products of
AGMs and odd moments
of 6 Bessel functions may be obtained
from the vanishing of the contour integral
$\oint_C\widetilde{D}^2(z)z\,{\rm d}z$. {}From its imaginary
part, I obtain the superconvergence relation
\begin{equation}
\Im\oint_C\widetilde{D}^2(z)z\,{\rm d}z
=\int_\frac13^1\rho(x)\sigma_1(x)x\,{\rm d}x
-2\int_1^\infty\rho(x)\sigma_2(x)x\,{\rm d}x=0
\label{D2-imag}
\end{equation}
and hence, by cubic modular transformation
of both $\sigma_2$ and $\rho$, the identity
\begin{equation}
\frac{1}{4\sqrt3\pi}
\int_\frac13^1\frac{D(y)D\left(\frac{1}{3y}\right)}{y}\,{\rm d}y
=\int_0^\frac13D(y)\widetilde{D}(y)\,{\rm d}y
=\int_0^\infty t\,I_0^3(t)K_0^3(t)\,{\rm d}t
\label{D2-Dt}
\end{equation}
where the Bessel moment was derived from
the integral of $D\widetilde{D}$ in~\cite[Eq.~223]{B3G}.

For the real part of the contour integral, I use the sesquiplicate
transformation~(\ref{a-to-b}) to prove that
\begin{eqnarray}
I_1\,:=\,\int_0^\infty E^2(w)w\,{\rm d}w
&=&\int_\frac13^1\frac{D^2(x)}{18x}\,{\rm d}x\,:=\,I_2
\label{I12}\\
I_3\,:=\,\pi^2\int_0^\frac13\widetilde{D}^2(y)\,{\rm d}y
&=&\int_0^\frac13\frac{D^2(x)}{6x}\,{\rm d}x\,:=\,I_4
\label{I34}
\end{eqnarray}
and make cubic transformations of $\sigma_2$ and $\rho$
to obtain
\begin{equation}
\pi^2\Re\oint_C\widetilde{D}^2(z)z\,{\rm d}z=
3(I_1-I_2)+2(I_4-I_3)=0
\label{contour-real}
\end{equation}
with no further relation obtained by contour integration.

Then I use the distribution~(\ref{J-J}) to prove that
\begin{equation}
I:=\int_0^\infty t\,I_0^2(t)K_0^4(t)\,{\rm d}t
=\,\int_0^\infty E^2(w)w\,{\rm d}w
\label{fold2}
\end{equation}
by folding two copies of~(\ref{E-moment-w}).
Next I remark that the appearance of $D$ as a moment,
in~(\ref{Dy-moment}), and also as a spectral function, in~(\ref{Dy-disp}),
proves that $I=I_4$, as was remarked in~\cite[Sect.~6.1]{B3G}.
Thus I obtain
\begin{equation}
I_1=I_2=I_3=I_4=I:=\int_0^\infty t\,I_0^2(t)K_0^4(t)\,{\rm d}t
\label{all-4}
\end{equation}
with each integral in~(\ref{I12},\ref{I34}) yielding the
same Bessel moment.

Finally, I show how to compute $\widetilde{D}(z)$
throughout the quadrant $z=x+{\rm i}y$ with $x>0$ and $y>0$,
using the separatrices
\begin{equation}
y_1(x)=\frac{x+1}{\sqrt{1+\frac{2}{3x}}}\,,\quad
y_2(x)=\frac{x-1}{\sqrt{1-\frac{2}{3x}}}
\label{sep}
\end{equation}
to distinguish the three cases
\begin{equation}
\widetilde{D}(z)=\left\{\begin{array}{rl}
2\rho(z)+\frac{1}{3}{\rm i}\sigma_2(z)&
\mbox{if }x>0\mbox{ and }y_1(x)<y\\
-2\rho(z)+{\rm i}\sigma_2(z)&
\mbox{if }x>1\mbox{ and }y_2(x)>y>0\\
{\rm i}\sigma_2(z)&
\mbox{otherwise}
\end{array}\right.
\label{3-cases}
\end{equation}
obtained by comparing results of numerical integration
in~(\ref{Dt-disp}) with the easier evaluations of complex AGMs
in~(\ref{sigma2}) and~(\ref{rho}) computed as convergents of the
defining iteration
${\rm AGM}(a,b)={\rm AGM}\left(\frac{a+b}{2},\sqrt{a b}\right)$.

\section{Conclusion}

In 1936, Wilfrid Norman Bailey proved an
identity~\cite[Eq.~3.3]{bailey1936} that leads,
via the analysis in~\cite{B3G,davdel,glasser,Joyce94},
to remarkable connections between Feynman diagrams,
integrals of the reciprocals of
arithmetic-geometric means, lattice Green functions and the
enumeration of staircase polygons. The cubic modular
transformation~(\ref{cubic-mod}) and the sesquiplicate modular
transformation~(\ref{a-to-b}) provide wonderful relations between
the integrals and generating functions for these four allied
structures.

\paragraph{Acknowledgements.} I am most grateful to David Bailey,
Gabriel Barton, Jonathan Borwein, Andrei Davydychev, Bob
Delbourgo, Larry Glasser, Geoffrey Joyce, St\'ephane Ouvry,
Neil Sloane, Bas Tausk and Jon Zucker, for advice and encouragement. 
I thank Herbert Gangl and Dirk Kreimer for inviting me to talk on 
singular values of elliptic integrals\footnote{Lecture recorded at {\tt
http://durpgap.googlepages.com/qftw}~.} at the workshop on {\sl
Hopf algebras and periods in quantum field theory} held in
Durham, in January 2008.

\raggedright

\end{document}